\documentclass[a4paper,11pt]{article}

\usepackage{authblk}   % package for authors & affiliations
\usepackage{natbib}
\usepackage[english]{babel}
\bibpunct{(}{)}{;}{a}{}{,}
\usepackage{epsfig}
\usepackage{fancyhdr,amssymb,a4wide,hyperref}
\hypersetup{
    colorlinks=true, %set true if you want colored links
    linktoc=all,     %set to all if you want both sections and subsections linked
    citecolor=blue,
}
\usepackage{amsmath}
\usepackage{graphicx}
\usepackage{tabularx}
\usepackage{multicol}
\usepackage[dvipsnames]{xcolor}
\usepackage[a4paper, top=1.6cm]{geometry}
\newcommand{\equalcontrib}{\textsuperscript{*}}

\title{Milky Way Disc \& Bulge in-situ populations \\ \normalsize ESO White Paper  Expanding Horizons call }

\author[1]{M.\,Bergemann\equalcontrib}
\author[2]{G.\,Kordopatis\equalcontrib}
\author[3,4]{G.\,Casali}
\author[5]{S.\,Khoperskov}
\author[6]{P.\,McMillan}
\author[5]{L.\,Marques}
\author[5]{I.\,Minchev}

\author[7]{E.\,Poggio} 
\author[1]{M. Schultheis}
\author[8]{C.\,Viscasillas\,Vázquez}
\author[9]{H.-F.\,Wang} 
\author[10]{V.\,Grisoni} 
\author[2]{V.\,Hill}
\author[11]{R.\,Smiljanic} 

\affil[1]{\footnotesize Max Planck Institute for Astronomy, Koenigstuhl 17, Heidelberg, 69117, Germany}
\affil[2]{\footnotesize Universit\'e C\^ote d’Azur, Observatoire de la C\^ote d’Azur, CNRS, Laboratoire Lagrange, Nice, France}
\affil[3]{\footnotesize Research School of Astronomy and Astrophysics, The Australian National University, Canberra, ACT 2611, Australia}
\affil[4]{\footnotesize INAF – Osservatorio di Astrofisica e Scienza dello Spazio, via P. Gobetti 93/3, 40129 Bologna, Italy}
\affil[5]{\footnotesize Leibniz-Institut für Astrophysik Potsdam (AIP), An der Sternwarte 16, 14482 Potsdam, Germany}
\affil[6]{\footnotesize School of Physics \& Astronomy, University of Leicester, University Road, Leicester LE1 7RH, UK}
\affil[7]{\footnotesize INAF - Osservatorio Astrofisico di Torino, via Osservatorio 20, 10025 Pino Torinese (TO), Italy}
\affil[8]{\footnotesize Institute of Theoretical Physics and Astronomy, Vilnius University, Sauletekio av. 3, 10257 Vilnius, Lithuania.}
\affil[9]{\footnotesize Dipartimento di Fisica e Astronomia “Galileo Galilei”, Università degli Studi di Padova, Vicolo Osservatorio 3, I-35122, Padova, Italy}
\affil[10]{\footnotesize INAF, Osservatorio Astronomico di Trieste, via G.B. Tiepolo 11, I-34131, Trieste, Italy}
\affil[11]{\footnotesize Nicolaus Copernicus Astronomical Center, Polish Academy of Sciences, Bartycka 18, 00-716 Warsaw, Poland}

\date{December 2025}

\begin{document}

\maketitle
% Add a single footnote manually for equal contribution
\begingroup
\renewcommand\thefootnote{*}
\footnotetext{These authors contributed equally to this work.}
\endgroup

\begin{abstract}
The formation and evolution of the Milky Way’s disc, bar, and bulge remain fundamentally limited by the lack of a contiguous, Galaxy-wide, high-precision chemo-dynamical map. Key open questions - including the survival or destruction of the primitive discs, the origin of the bulge’s multi-component structure, the role of mergers and secular processes, and the coupling between stellar chemistry, dynamics, and the Galactic potential — cannot be fully resolved with current or planned facilities. Existing spectroscopic surveys provide either high resolution for small samples or wide coverage at insufficient resolution and depth, and none can obtain homogeneous abundances,
%($\leq0.05$\,dex),
precise 3D kinematics, and reliable ages for the millions of stars required, particularly in the obscured midplane, the far-side of the bar, or the outer, low-density disc. A new wide-field, massively multiplexed, large-aperture spectroscopic facility, capable of both high- and low-resolution spectroscopy over tens of thousands of square degrees, is therefore essential. Such a facility would deliver the statistical power, sensitivity, and completeness needed to reconstruct the Galaxy’s assembly history, constrain its gravitational potential, and establish the Milky Way as the definitive benchmark for galaxy evolution.

\end{abstract}

\newpage

%[Figure 1 placeholder: Schematic of the Milky Way disc and bulge with regions requiring uniform spectroscopic coverage highlighted]

%\bigskip 

\section{Science case}

Over the past decade, advances in the volume and quality of stellar data, together with increasingly realistic galaxy formation models, have driven a paradigm shift in our understanding of the formation of the Galactic disc and bulge. Proposed evolutionary pathways span a wide range of scenarios, from sequential assembly of Galactic components \citep{Belokurov22, XiangRix25}, to complex high-redshift disc formation shaped by mergers and accretion \citep{Sotillo-Ramos23, Semenov25}, and the transformation of early discs into pseudo-bulges \citep[e.g.][]{vanDonkelaar25}. At the same time, observations are beginning to reveal the chemo-dynamical regimes in which these models can be tested. 

The discovery of very metal-poor stars on dynamically cold orbits \citep[e.g.][]{Sestito20, Gent24, Nepal24}, along with the  complexity of in-situ stellar populations of the bulge \citep{Bensby2017, Horta2025}, draws striking parallels with Population\,III candidates identified in high-redshift galaxies \citep{Visbal25}. This emerging body of evidence places renewed emphasis on the very- and ultra-metal-poor regimes of the disc and bulge, while also opening new avenues for directing searches for surviving Population\,III stars \citep{Magg18, Klessen23}.
Although forthcoming spectroscopic surveys such as 4MOST \citep{deJong19} and WEAVE \citep{Jin24} will efficiently follow up metal-poor candidates and greatly expand the available samples, they are likely to lack the combination of sample size and spectral resolution required to robustly determine the fundamental properties of these populations, such as their total mass, metallicity floor, formation epochs, and role in the earliest phases of Galactic evolution \citep[e.g.][]{Gonzalez-Rivera24}.

A comprehensive chemo-dynamic map is also essential for understanding the role of secular processes in the disc evolution and for breaking down the degeneracies associated with the non-equilibrium state of the Galaxy. Ridges, arches, streaming motions and corrugations observed in the space of kinematics and chemical abundances suggest an intricate dynamical interplay between the bar, spiral arms, the warp, and external perturbers such as the Sagittarius dwarf galaxy, the Large Magellanic Cloud, or the Gaia-Sausage Enceladus \citep[e.g.][]{minchev09, Antoja18, Bergemann18, GaiaKatz18, Carrillo18, Laporte19, Poggio22, Hawkins23, Khoperskov23a, Wang24,  Vislosky24, Kordopatis25}. Reconstructing the sequence and strength of these interactions requires both precise elemental abundances ($\delta$[X/Fe]$\lesssim$0.05\,dex) and accurate 3D kinematics for stars distributed over the entire disc, including its outermost sparsely-populated flared regions, the heavily dust-obscured midplane, and the far-end of the bar \citep[revealing whether asymmetric interactions are responsible for non-synchronous star formation and radial migration patterns across the Galaxy, e.g.][]{Marques25}. No facility available or planned for the 2030s can provide the combination of sample size and sensitivity to probe the entire Galactic volume.

The Galactic bulge presents a comparable set of challenges. It hosts a remarkably wide range of stellar ages and metallicities \citep[see the review of ][]{Zoccali24}, with populations that may originate from the early disc, and disrupted dwarf galaxies and clusters \citep{Barbuy18}. The inner kiloparsec also contain the nuclear stellar disc and the nuclear star cluster, each with distinct kinematic and chemical signatures that remain only partially understood \citep{Schultheis25}. Upcoming surveys (such as 4MOST or MOONS GTO, \citealt{Gonzalez20}) will provide only incomplete coverage of these regions due to extinction, crowding,  the lack of sufficient spectroscopic depth or insufficiently large aperture. As a result, the precise mapping of the bulge’s structural components, their origins, and their interrelations  will remain beyond reach.

Reconstructing the star-formation history of the entire Milky Way further requires the ability to determine the birth radii and chemical evolution pathways of stars throughout the disc and bulge. Because radial migration decouples present-day stellar positions from their birth environments \citep[e.g.][]{Sellwood02, Schonrich09a, Minchev12c, Roskar12, Kordopatis15b, ViscasillasVazquez23}, achieving this goal demands precise age estimates and detailed abundances (including light elements and neutron-capture species) for very large stellar samples \citep{Casali25}. Only with millions of such measurements can we assemble a self-consistent picture of how the Milky Way enriched its gas reservoirs, built its disc, and sustained star formation across cosmic time.

A similarly fundamental objective is the determination of the Galactic gravitational potential in the baryon-dominated regime. High-resolution spectra spanning multiple nucleosynthetic families  will tie stellar chemistry to orbital families and Galactic dynamics, and therefore the gravitational potential and dark matter density of the Milky Way, and disentangle in-situ versus accreted populations \citep[e.g.][]{Horta24}. Achieving this at the necessary precision requires homogeneous, Galaxy-wide, $R\gtrsim40,000$ spectroscopy, again exceeding the capabilities of any expected survey in the 2030s.

Finally, establishing the Milky Way as a resolved benchmark for galaxy evolution requires datasets comparable in fidelity to extragalactic integral-field surveys (e.g. MANGA, SAMI, \citealt{Bundy15, Bryant15}) revealing the detailed kinematics and stellar populations properties across thousands of galaxies in the nearby universe. At the same time JWST and ALMA are revealing the existence of disc galaxies at redshifts  $z\gtrsim$ 4-6 \citep{Fraternali21, Roman-Oliveira23, Jacobs23}, and it is now essential to decipher the complete chemo-dynamical structure of our own Galaxy in order to understand how such early discs formed and evolved. Present and upcoming facilities fall short of enabling this comparison at the required completeness and precision.

\section{Why is a new facility required beyond 2030s}
Although the next decade will bring powerful instruments (4MOST, WEAVE, MOONS, SDSS-V, LSST, Roman, PLATO, and possibly Gaia-NIR) none will provide the unified, deep, contiguous, high-resolution spectroscopic dataset required to answer the questions outlined above. Their limitations are structural rather than incremental: they lack either the aperture, the multiplexing capacity, the resolution, the wavelength coverage, or the ability to observe the heavily reddened regions of the Galactic plane and bulge.
Most importantly, they cannot deliver high-resolution spectroscopy (R $\sim$ 40,000) for tens of millions of stars, nor can they provide spatially complete coverage of the disc midplane,  the inner bulge or the disc outskirts at different azimuths and over a volume of 20\,kpc wide. The small fields of view of ELT-class instruments \citep[MOSAIC,][]{Evans15}, combined with their lack of multiplexing \citep[ANDES,][]{Marconi24}, make them unsuitable for the large-scale surveys required. Even with the combined strengths of all planned facilities, the Galaxy-wide, homogeneous abundance map that is needed to constrain early disc formation, bulge structure, and radial migration will remain out of reach in the 2030s.

\section{Facility requirements}
To address the science questions described above, a future facility must integrate  a 10m-class collecting area to reach the faint ($m_G \gtrsim 20$), distant and/or heavily reddened stellar populations most relevant for reconstructing the Galaxy’s early phases and current in-plane dynamics. It must also support both high-resolution ($R\gtrsim40,000$) and low-resolution ($\gtrsim 5,000$) spectroscopy over a wide field of view, enabling simultaneous precise chemical abundance measurements \citep{Kordopatis23b} %\footnote{see \url{https:line-detector.oca.eu} and \citet{Kordopatis23b}.}
and efficient  candidate selection for all stellar populations, including rare stellar types such as (very/ultra) metal-poor stars. Third, it must deploy thousands of high-resolution fibres and tens of thousands of low-resolution fibres to perform a truly comprehensive and efficient Galactic survey over a relatively short period (e.g. 5 years). Finally, it must provide homogeneous, contiguous coverage across more than 30,000 square degrees, including the midplane, warp, bulge, and bar.
Such a facility would uniquely enable a uniform, precise abundance catalogue (including n-capture, iron-peak, and light elements) with precisions better than 0.05\,dex for tens of millions of stars. It would also provide the radial velocities and parameter accuracy required to exploit the full potential of Gaia, LSST, PLATO, Roman, and future astrometric missions (Gaia-NIR).

\section{Technology and data-handling requirements }
The technological demands of this facility rely on evolutions of existing instrumentation rather than revolutionary breakthroughs: fibre positioners capable of placing few thousand high-resolution fibres across a wide field,  high-resolution spectrographs (four or five windows, from the blue to the near IR), low-resolution spectrographs covering the full optical range down to faint magnitudes (required for turnoff, metal-poor stars, and/or reddenned targets). A high-resolution infrared module would further enhance coverage of obscured bulge regions and maximize synergy with Gaia-NIR and Roman.

The primary challenge lies in data handling. Processing tens of millions of spectra requires automated, scalable pipelines for reduction, atmospheric-parameter inference, and abundance determination. The resulting catalogues will be petabyte-scale but can be served easily either through  virtual-observatory standards and/or cloud-based architectures.

\section{Conclusion}
To achieve the objectives outlined above, a wide-field, massively multiplexed, 10m-class spectroscopic facility would be required. With such a facility, the community would have, for the first time, a complete and precise chemo-dynamical map of the Milky Way’s disc, bulge, and bar. This dataset would reveal the nature of the earliest in-situ populations, reconstruct the sequence of dynamical interactions shaping the disc, characterize the formation pathways of the bulge, and determine the Galactic gravitational potential with unprecedented accuracy. It would also provide the benchmark needed to interpret low- and high-redshift extragalactic  observations and connect the Milky Way to the broader context of galaxy evolution.

\begin{multicols}{3}
\tiny{
\bibliographystyle{aa_gk}
\bibliography{Bibliography}
}
\end{multicols}
\end{document}